\documentclass[fleqn,usenatbib]{mnras}

\usepackage{newtxtext,newtxmath}

\usepackage[T1]{fontenc}

\DeclareRobustCommand{\VAN}[3]{#2}
\let\VANthebibliography\thebibliography
\def\thebibliography{\DeclareRobustCommand{\VAN}[3]{##3}\VANthebibliography}

\usepackage{graphicx}	
\usepackage{amsmath}	

\newcommand{\fermi}{\textit{Fermi}}
\newcommand{\blinky}{{4FGL}~J1646}
\newcommand{\blinkyaskap}{{ASKAP}~J1646}

\newcommand{\periodh}{$5.267104\pm{3\times 10^{-6}}$}

\newcommand{\periodhval}{$5.267104$}

\title[Radio eclipses from 4FGL~J1646.5$-$4406]{Discovery of radio eclipses from {4FGL}~J1646.5$-$4406: a new candidate redback pulsar binary}

\author[A. Zic, Z. Wang et al.]{Andrew Zic,$^{1}$\thanks{E-mail: andrew.zic@csiro.au}
Ziteng Wang,$^{2,3}$\thanks{E-mail: ziteng.wang@curtin.edu.au}
{Emil Lenc,$^{1}$}
{David L. Kaplan,$^{4}$}
{Tara Murphy,$^{2,5}$} 
{A. Ridolfi,$^{6,7}$}
\newauthor
{Rahul Sengar,$^{4}$} 
{Natasha Hurley-Walker,$^{3}$}
{Dougal Dobie,$^{8,5}$}
{James K.~Leung,$^{2,1,5}$}
{Joshua~Pritchard,$^{2,1,5}$}
\newauthor
and {Yuanming Wang$^{8,5}$}
\\
$^{1}$Australia Telescope National Facility, CSIRO, Space and Astronomy, PO Box 76, Epping, NSW 1710, Australia\\
$^{2}$Sydney Institute for Astronomy, School of Physics, University of Sydney, Sydney, NSW 2006, Australa\\
$^{3}$International Centre for Radio Astronomy Research, Curtin University, Bentley, WA 6102, Australia\\
$^{4}$Center for Gravitation, Cosmology, and Astrophysics, Department of Physics, University of Wisconsin-Milwaukee, P.O. Box 413, Milwaukee, WI 53201, USA\\
$^{5}$ARC Centre of Excellence for Gravitational Wave Discovery (OzGrav), Hawthorn, Victoria, Australia\\
$^{6}$INAF – Osservatorio Astronomico di Cagliari, Via della Scienza 5, I-09047 Selargius (CA), Italy\\
$^{7}$Max-Planck Institut fur Radioastronomie, Auf dem H\"ugel 69, D-53121 Bonn, Germany\\
$^{8}$Centre for Astrophysics and Supercomputing, Swinburne University of Technology, Hawthorn, Victoria, Australia\\
}

\date{Accepted XXX. Received YYY; in original form ZZZ}

\pubyear{2023}

\begin{document}
\label{firstpage}
\pagerange{\pageref{firstpage}--\pageref{lastpage}}
\maketitle

\begin{abstract}
Large widefield surveys make possible the serendipitous discovery of rare sub-classes of pulsars. One such class are ``spider''-type pulsar binaries, comprised of a pulsar in a compact orbit with a low-mass (sub)stellar companion. In a search for circularly-polarized radio sources in ASKAP Pilot Survey observations, we discovered highly variable and circularly polarized emission from a radio source within the error region of the $\gamma$-ray source {4FGL}~J1646.5$-$4406. The variability is consistent with the eclipse of a compact, steep-spectrum source behind ablated material from a companion in a $\sim 5.3$\,h binary orbit. Based on the eclipse properties and spatial coincidence with {4FGL}~J1646.5$-$4406, we argue that the source is likely a recycled pulsar in a ``redback'' binary system. Using properties of the eclipses from ASKAP and Murchison Widefield Array observations, we provide broad constraints on the properties of the eclipse medium. We identified a potential optical/infra-red counterpart in archival data consistent with a variable low-mass star. Using the Parkes Radio Telescope ``Murriyang'' and MeerKAT, we searched extensively for radio pulsations but yielded no viable detections of pulsed emission. We suggest that the non-detection of pulses is due to scattering in the intra-binary material, but scattering from the ISM can also plausibly explain the pulse non-detections if the interstellar dispersion measure exceeds $\sim$600\,pc\,cm$^{-3}$. Orbital constraints derived from optical observations of the counterpart would be highly valuable for future $\gamma$-ray pulsation searches, which may confirm the source nature as a pulsar.
\end{abstract}




\section{Introduction}
\label{sec:intro}
Millisecond pulsars (MSPs; periods $<30$\,ms), are excellent tools for a range of astrophysical applications, including stringent tests of general relativity \citep{2021PhRvX..11d1050K}, studying the equation-of-state for neutron degenerate matter \citep{2019ApJ...887L..24M}, searches for nanohertz-frequency gravitational waves \citep{2023RAA....23g5024X,2023ApJ...951L...6R,2023A&A...678A..50E,2023ApJ...951L...8A}, and studying binary evolution \citep{2012ApJ...753L..33B,2015MNRAS.446.2540S,2013ApJ...775...27C}. They are formed after accreting matter from a companion star and spinning up to millisecond spin periods \citep{1982Natur.300..728A}, after which they are typically found in binary orbits with periods $P_b\gtrsim 0.5$\,d with their donor companions \citep{2006csxs.book..623T}, though several isolated millisecond pulsars also exist.

``Spiders" are a rare class of MSP binary systems in which material from a low-mass companion in a compact orbit ($P_b < 1$\,d) is ablated by the relativistic pulsar wind \citep{2013IAUS..291..127R}. Spiders are comprised of two sub-classes, known as ``black widows'' with degenerate companions with masses $\sim 0.01-0.05 \,M_\odot$, and ``redbacks'' with main sequence companions with masses $>0.1\,M_\odot$ \citep{2013IAUS..291..127R}. The ablated companion material often causes radio eclipses, which occur due to enhanced dispersion, scattering, or absorption \citep{1988Natur.333..237F,1996ApJ...465L.119S,2020MNRAS.494.2948P,2021ApJ...922L..13W, 2022MNRAS.513.1794B}. 

Spiders likely represent very early stages in the life-cycle of recycled pulsars \citep{2014ApJ...786L...7B}, after the neutron star has completed accretion from the donor star. This is supported by the observation of multiple ``transitional'' MSPs, which switch between a radio-pulsar mode and a Low-mass X-ray binary (LMXB) mode, where mass transfer occurs \citep[e.g.][]{2009Sci...324.1411A,2015ApJ...800L..12R}. There is both theoretical \citep{2014ApJ...786L...7B} and observational \citep{2019ApJ...872...42S,2023NatAs...7..451C} support for the mass of spider pulsars being systematically higher than normal pulsar masses, due to the recent completion of accretion from the donor star, making these systems particularly useful for studies of the neutron star equation of state \citep{2016ARA&A..54..401O}.   
It also seems apparent that there is an evolutionary link between LMXBs, redbacks, black widows, and field MSPs, but multiple formation mechanisms for redback and black widow systems have been proposed \citep{2013ApJ...775...27C,2015MNRAS.446.2540S,2020MNRAS.493.2171D,2021MNRAS.500.1592G}. Discovery of new and unusual spiders may provide useful insights into the demographics and formation pathways of these systems \citep{2023Natur.620..961P}. 

The Large-Area Telescope on the \textit{Fermi} Space Telescope \citep[\textit{Fermi}-LAT;][]{2009ApJ...697.1071A} has enabled a wealth of pulsar discoveries via follow-up of unassociated $\gamma$-ray sources. This has included the majority of spider-type systems \citep[e.g., ][]{2013MNRAS.429.1633B,2013ApJ...773L..12B,2015ApJ...810...85C,2016ApJ...819...34C,2023MNRAS.519.5590C, 2023ApJ...943..103A}. There is also growing population of convincing spider candidates (lacking detections of pulses at radio or $\gamma$-ray wavelengths) identified by follow-up of \textit{Fermi} sources at X-ray and optical wavelengths \citep[e.g.][]{2014ApJ...788L..27S, 2015ApJ...813L..26S, 2015ApJ...812L..24R, 2020MNRAS.497.5364B, 2022ApJ...935....2C, 2022Natur.605...41B}.
Searches for pulsar candidates in radio continuum imaging datasets is also an increasingly successful approach. These include identification of radio sources exhibiting a steep radio spectrum \citep{1982Natur.300..615B, 2018MNRAS.475..942F, 2022ApJ...927..216R}, scintillation \citep{2016MNRAS.462.3115D,2017MNRAS.472.1458D}, variability \citep{2022ApJ...930...38W}, high degrees of polarization \citep{2019ApJ...884...96K, 2022A&A...661A..87S}, globular cluster association \citep{2023MNRAS.525L..76H}, and association with \textit{Fermi} sources \citep{2018MNRAS.475..942F}.



In this paper, we report our analysis and multi-wavelength follow up following new detections of the putative radio counterpart to the $\gamma$-ray source \textsc{4FGL}~J1646.5$-$4406 (hereafter \blinky{}).
The source was previously investigated by \citet{2018MNRAS.475..942F} owing to its positional correspondence with the unassociated \textit{Fermi} source \blinky{}, its apparent angular compactness, and its steep radio spectrum -- properties consistent with a pulsar origin. However, \citet{2018MNRAS.475..942F} determined it was unlikely to be a pulsar due to a high ratio of its integrated to peak radio flux density, which they proposed to be due to the source being partially resolved. In this work, we revisit the nature of this source after we detected highly polarized emission and periodic radio variability, which we ascribe to the eclipse of a pulsar behind material from a companion. We summarise our unsuccessful search for pulsed emission from this source, but argue that the source is nonetheless most likely a radio pulsar in a redback binary system. We discuss possible reasons for the non-detection of pulses and place initial constraints on the eclipse mechanism and properties of the eclipsing medium.

\section{Observations and analysis} \label{sec:obs}

\begin{table*}
    \centering
    \begin{tabular}{ccccccc}
    \hline
    {Telescope }  & {Start} & {$\Delta t$} & {$\nu$} & {$\Delta \nu$} & {Mode} & $\phi$ \\
    {}        & {(UT)}                      & {(h)}      & {(MHz)}             & {(MHz)} & {} & {} \\
    \hline
    \hline
         ASKAP     & 2021~Jun~26 09:22:55 & 9.9 & 943 & 288 & Imaging   &  0 -- 1 \\
         ASKAP     & 2021-Sep~19 19:16:16 & 1.0 & 943 & 288 & Imaging   &  0.59 -- 0.78 \\
         ATCA      & 2021~Sep~21 01:12:55 & 3.6 & 2100 & 2048 & Imaging   & 0.28 -- 0.09 \\
         Murriyang & 2021~Sep~21 03:41:40 & 1.0 & 2368 & 3328 & Pulsar search  &  0.75 -- 0.94 \\
         ASKAP     & 2021~Sep~20 12:54:53 & 9.8 & 943 & 288 & Imaging   &  0 -- 1 \\
         ASKAP     & 2022~Feb~25 22:24:40 & 5.5 & 943 & 288 & Imaging   &  0 -- 1 \\
         MeerKAT   & 2022~Mar~07 03:39:39 & 0.67 & 1284 & 856 & Imaging$+$Pulsar search  & 0.69 -- 0.82 \\
         MeerKAT   & 2022~Mar~09 03:01:15 & 0.67 & 1284 & 856 & Imaging+Pulsar search  & 0.68 -- 0.81 \\
         MeerKAT   & 2022~Mar~19 00:00:29 & 0.67 & 1284 & 856 & Imaging$+$Pulsar search  & 0.68 -- 0.80 \\
         MeerKAT   & 2022~Mar~20 23:24:56 & 0.67 & 1284 & 856 & Pulsar search  & 0.68 -- 0.80 \\
         MWA & 2022~June~02 -- September~22 & $21\times1.5$ & 200 & 30 & Imaging  &  0 -- 1 \\
         \hline
    \end{tabular}
    \caption{Summary of radio observations. Notes: $\Delta t$ is the integration time on-source, and $\Delta \nu$ is the total bandwidth around central frequency $\nu$. $\phi$ is the inferred orbital phase spanned by the observations. For the MWA, we summarise the twenty-one observations taken from the Galactic Plane Monitoring program used for this source.\label{tab:obs_table}}
\end{table*}


\subsection{Australian SKA Pathfinder (ASKAP)}\label{subsec:ASKAP_obs}

\begin{figure*}
    \centering
    \includegraphics[width=\linewidth]{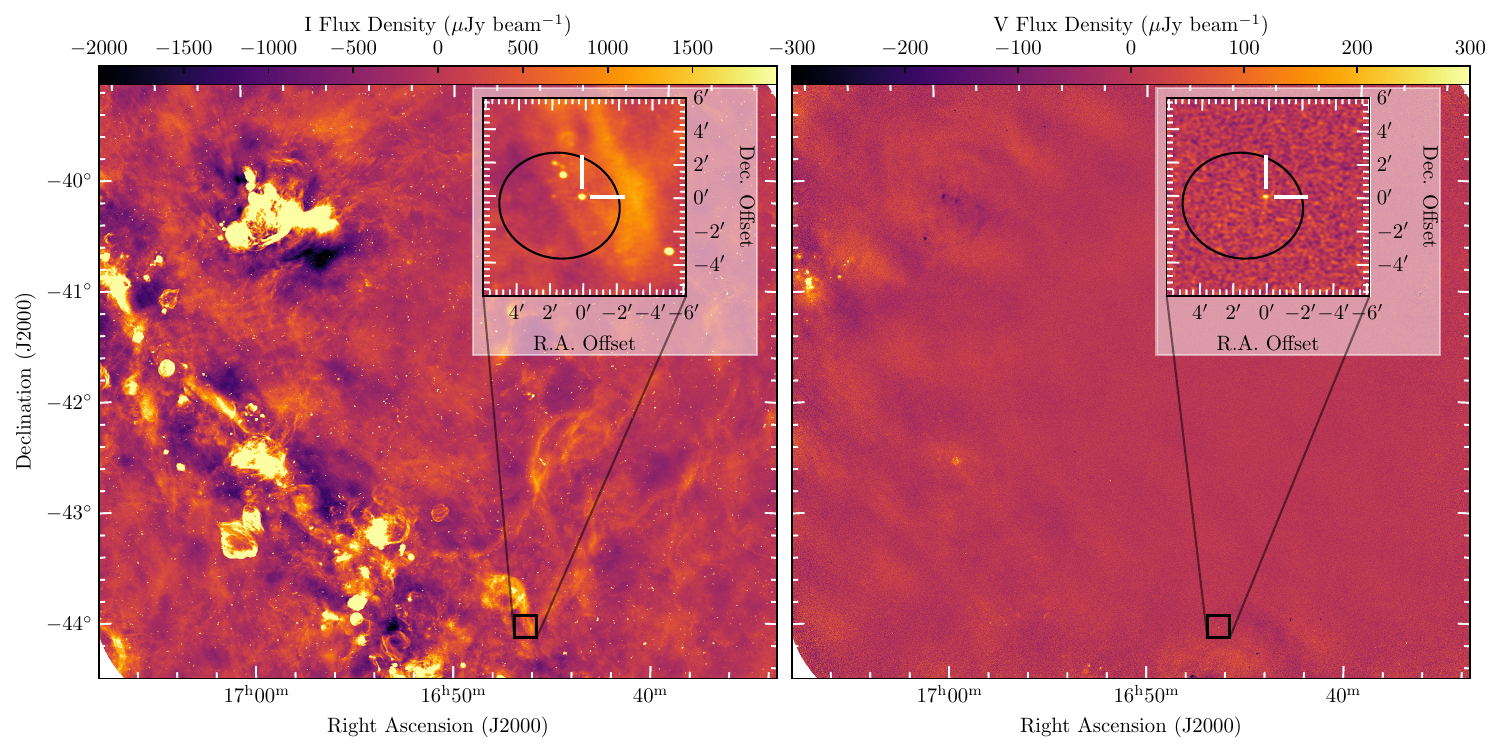}
    \caption{\label{fig:field_image}ASKAP field image in Stokes $I$ (left) and $V$ (right). Inset on both axes is a $6\arcmin \times 6\arcmin$ cutout around \blinkyaskap{}, where the \textit{Fermi} 95\,per\,cent positional error ellipse around \blinky{} is shown in black. Note that direction-dependent leakage calibration has not been applied to the Stokes $V$ image, so leakage on the order of $\lesssim 1$\,per\,cent is visible toward bright, extended regions of emission.}
\end{figure*}

\begin{figure*}
    \centering
    \includegraphics[width=\linewidth]{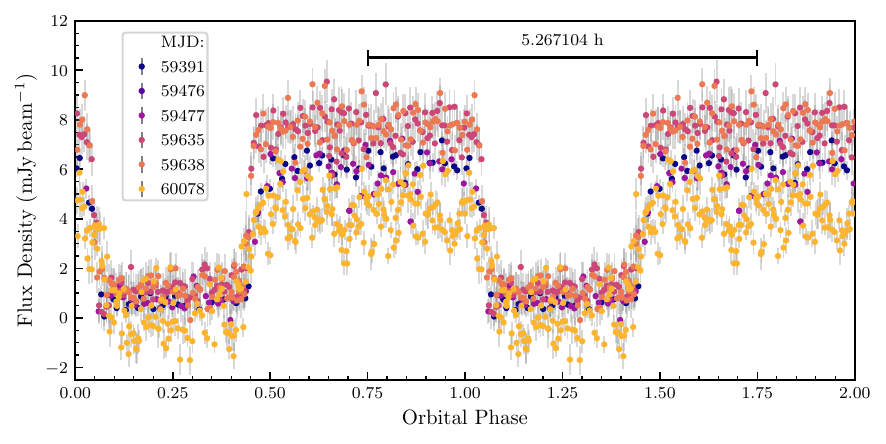}
    \caption{ASKAP radio lightcurve for \blinkyaskap{}, folded to the putative orbital period of \periodhval\,h. Points are coloured by their observing epoch. For clarity we repeat the data from phases 1--2 and provide a scale-bar to indicate the time-span of the orbit. {Long-term variability is evident through the differing out-of-eclipse flux densities across different epochs, ranging from $\sim 4$ -- 9\,mJy.}}
    \label{fig:eclipse_lc}
\end{figure*}

\begin{figure*}
    \centering
    \includegraphics[width=\linewidth]{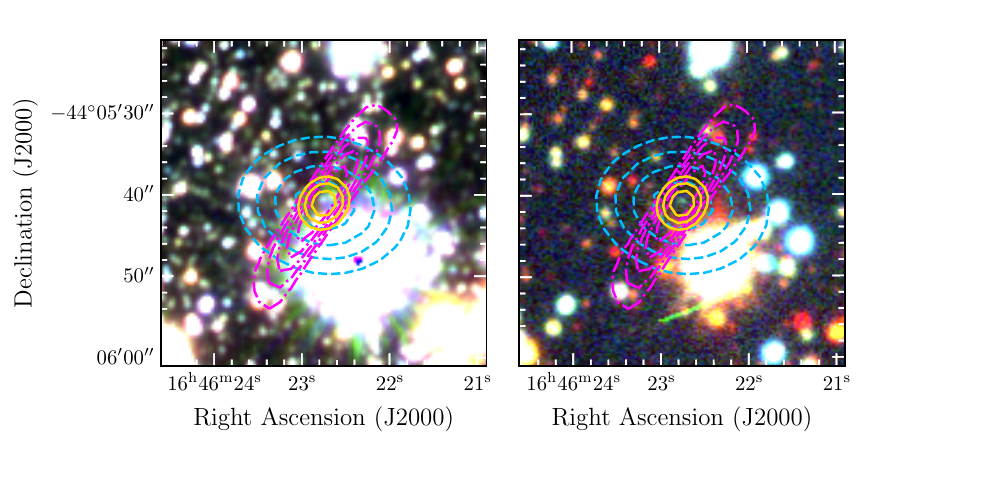}
    \caption{VVV $JHK_s$ (left) and DECaPS $grz$ (right) RGB-coloured cutouts around the position of \blinkyaskap{}, with radio contours from ASKAP {(blue, dashed)} and ATCA {(pink, dot-dashed)} and MeerKAT {(gold, solid)}.}
    \label{fig:radio_overlay}
\end{figure*}

We detected circularly-polarized emission from the radio source ASKAP~J164622$-$440540 (hereafter \blinkyaskap{}) in an observation taken for the ASKAP EMU Pilot Survey \citep{2021PASA...38...46N} (Schedule Block ID 28280, observed 2021 June 26). The observations were taken at a central frequency of 943.41\,MHz, with 36 beams tiled in a close-pack configuration. The central field coordinates were RA=$16^{\text{h}}50^{\text{m}}46^{\text{s}}$ Dec=$-41^{\circ}52^{\prime}44\arcsec$. The data were calibrated using ASKAPsoft version 1.2.1 and CASA version 6.2. Further details on processing and imaging can be found in \citet{2021PASA...38...46N}. We show the observation field in Stokes $I$ and $V$ in Fig. \ref{fig:field_image}. 

We detected \blinkyaskap{} as a point source in both Stokes $I$ and $V$, with flux densities of $5.19\pm0.06$\,mJy and $+0.38\pm0.03$\,mJy respectively.\footnote{In this paper, we use the IAU definition of Stokes $V$, where positive (negative) Stokes $V$ corresponds to right (left) handed circular polarization.} The resulting polarization fraction $f_c=7.3\pm0.6$\,per\,cent. This is much higher than the typical circular polarization leakage fraction of $\sim -0.4$\,per\,cent in the local region around \blinkyaskap{}, determined by inspection of nearby bright Stokes $I$ sources.

Following the procedure in \citet{2019MNRAS.488..559Z}, we constructed lightcurves and dynamic spectra for the source by subtracting the sky model constructed during the deconvolution process from the calibrated visibilities. We then phase-rotated the visibilities to the target position and vector-averaged all baselines, yielding complex visibility amplitudes for each correlator integration cycle, frequency channel, and instrumental polarization. We transformed the data from the native instrumental polarization basis ($XX$, $YY$, $XY$, $YX$) into the Stokes basis ($I$, $Q$, $U$, $V$), and produced lightcurves by averaging the data across frequency.

We found strong, eclipse-like variability in the initial EMU pilot observation, with bright ($S_{943}\sim 7.5$\,mJy), stable emission present for approximately 3 hours, followed by a period of no detectable emission lasting approximately 2 hours. This behaviour was then repeated. In order to confirm the periodic nature of the variability, we conducted further observations with ASKAP on 2021~September~20, 2022~February~25, and 2022~February~28, with identical observing configuration and processing methods to the original EMU pilot observation. After transforming the lightcurve timestamps to the frame of the solar system barycenter, we measured the period by fitting a double Fermi-Dirac function with a freely-inferred period and reference epoch (see \S \ref{subsec:ecl_props}). This fit yielded a period of \periodh{}\,h, with a reference epoch of MJD~$59391.3578\pm0.0005$. We show the ASKAP time-series from all observations folded at the measured period in Fig. \ref{fig:eclipse_lc}. 

The apparent non-eclipsed continuum flux density is comparable to measurements reported by \citet{2018MNRAS.475..942F}. In addition, the strong variability introduces point-spread function (PSF) artefacts in the continuum image around \blinkyaskap{}. These effects may be responsible for the inflated integrated-to-peak flux ratio reported for this radio source in \citet{2018MNRAS.475..942F}.



\subsection{Parkes/Murriyang pulsar search}
Motivated by the detection of periodic, eclipse-like variability and circular polarization from \blinkyaskap{}, along with the spatial coincidence with the $\gamma$-ray source \blinky{} \citep{2018MNRAS.475..942F}, we carried out pulsar search-mode observations with Parkes 64m Radio Telescope, Murriyang, on 2021 September 21 using Director's Discretionary Time (project code: PX080). The observations were carried out with the Ultra-Wideband Low (UWL) receiver \citep{2020PASA...37...12H}, which provides continuous frequency coverage spanning 704 to 4032\,MHz. We coordinated simultaneous observations with the Australia Telescope Compact Array (ATCA; see \S\ref{subsec:ATCA_obs}) to ensure that the emission was present during the pulsar search observations.


We performed a pulsar search using \textsc{presto}, to a maximum dispersion measure of 1800\,pc\,cm$^{-3}$ (corresponding to a distance $>25$\,kpc based on the YMW16 electron-density model \citep{2017ApJ...835...29Y}) and acceleration parameter $z_{\text{max}}=200$ (corresponding an orbital period of $\sim$5.4\,hr assuming a 3.5\,ms spin period and 20\,min integration). Since the radio source is quite bright at low frequencies (5.19\,mJy at 943\,MHz), we split the UWL band into six segments ($704$\,--\,$1344$\,MHz; $1344$\,--\,$1856$\,MHz; $1856$\,--\,$2368$\,MHz; $2368$\,--\,$2880$\,MHz; $2880$\,--\,$3520$\,MHz; $3520$\,--\,$4032$\,MHz) and searched each segment independently. Searches over the low-frequency segments allowed us to leverage the steep radio spectrum, while searches over the high-frequency segments mitigated potential pulse smearing due to high degrees of scattering (scattering timescale $\tau_s \sim \nu^{-4}$).

After manual inspection of all candidates above a S/N ratio of 4, we did not find any convincing pulsations. We describe further (unsuccessful) pulsar searches with MeerKAT in \S\ref{subsec:mkt_psrsearch} and discuss implications of non-detections in Section \ref{sec:discussion}.

\subsection{Australia Telescope Compact Array}
\label{subsec:ATCA_obs}
We took observations with the Australia Telescope Compact Array under Director's Discretionary Time (project code: CX489) with the 16\,cm receiver. We used the Compact Array Broadband Backend (CABB), which delivers $\sim$2\,GHz bandwidth centered at 2.1\,GHz. The observations were recorded with 1\,MHz channels and a correlator integration time of 10\,s. We observed PKS\,B1934$-$638 for approximately 10\,mins for flux density, bandpass, and polarization leakage calibration. We observed \blinkyaskap{} in 20\,min scans, book-ended between 2\,min scans on a nearby gain calibrator, J1646$-$50. Due to telescope maintenance at the time of observations, antenna CA02 was not available for observing. We flagged and calibrated the data following standard procedures with \textsc{miriad}. Owing to the steep spectrum of the source, we split the 2\,GHz band into two 1\,GHz segments centered at 1.588\,GHz and 2.612\,GHz. We used the \textsc{miriad} task \textsc{mfclean} to image each frequency segment. We used a briggs weighting with a robustness of 0.5, and deconvolved to a depth of $150$\,$\mu$Jy, corresponding to approximately $3\times$ the image rms. 

We detect the source with a flux density of $1.35\pm0.06$\,mJy at 1588\,MHz and $0.47\pm0.06$\,mJy at 2612\,MHz, yielding a spectral index $\alpha = -2.1\pm0.25$ ($S_\nu\propto \nu^{\alpha}$). We constructed lightcurves by subtracting the sky model generated through deconvolution and self-calibration from the calibrated visibilities using \textsc{uvmodel}, and fitted the visibilities with a point source model at the location of \blinkyaskap{} in 5- and 20-minute time intervals. We detected variability in the 20-minute lightcurves at 1588\,MHz consistent with eclipse egress near the beginning of the observation (the orbital phase spanned 0.28 -- 0.09), with the source not detected during the first 40 minutes. However, the signal-to-noise and time resolution of the lightcurves was not high enough for further modelling, and the source was too faint to detect significant variability at 2612\,MHz on relevant timescales. 

\subsection{MeerKAT Imaging}

We observed \blinkyaskap{} (see Table~\ref{tab:obs_table}) for 40\,min on March 7th, 9th, 19th, and 20th 2022 using the MeerKAT radio telescope at L-band, with a central frequency of 1.28\,GHz (856\,MHz bandwidth, 4096 channels) and an integration time of 8s (project code DDT-20220227-TM-01). Continuum imaging data were only recorded for observations on March 7, 9, and 19, whereas high time-resolution pulsar search data were recorded in all four observations (see \S \ref{subsec:mkt_psrsearch}). 
Each observation was scheduled to occur around superior conjunction of the pulsar using an initial eclipse ephemeris derived from ASKAP observations. The resulting continuum images had an rms noise of 20\,$\mu$Jy beam$^{-1}$. Both interferometric imaging data and pulsar search-mode data were recorded simultaneously in all MeerKAT observations. 
We used {PKS~B1934$-$638} for bandpass, flux density scale calibration, and {PKS~J1744$-$5144} for phase calibration. 
The archived raw visibilities were converted to Measurement Set format with the KAT Data Access Library ({\sc katdal}\footnote{\url{https://github.com/ska-sa/katdal}}) for further continuum imaging. 
We reduced the data using {\sc Oxkat}\footnote{\url{https://github.com/IanHeywood/oxkat}} \citep[v0.3;][]{2020ascl.soft09003H}, where the Common Astronomy Software Applications \citep[CASA;][]{2007ASPC..376..127M} package and {\sc Tricolour}\footnote{\url{https://github.com/ska-sa/tricolour}} \citep{2022ASPC..532..541H} were used for measurement sets splitting, cross calibration, flagging, {\sc CubiCal}\footnote{\url{https://github.com/ratt-ru/CubiCal}} \citep{2018MNRAS.478.2399K} was used for self-calibration, and {\sc Wsclean} \citep{2014MNRAS.444..606O} was used for continuum imaging. All processes were executed with {\sc Oxkat} L-band default settings. 

We measured peak flux densities of $3.42\pm0.02$\,mJy\,beam$^{-1}$ (Mar 07),  and $3.49\pm0.03$\,mJy\,beam$^{-1}$ (Mar 19) for \blinkyaskap{}. 
The best-fit position of the source based on the MeerKAT observations is: (J2000) $16^{\rm h}46^{\rm m}22\fs75\pm0\fs14~-44\degr05\arcmin41\farcs00\pm0\farcs14$.

We measured spectral, temporal, and polarization properties for \blinkyaskap{} with further reduction on two observations from March 7 and 19. 
The source had a steep radio spectrum within the bandpass in our MeerKAT observations, with spectral indices $\alpha=-3.1\pm0.1$ in both observations. The discrepancy between this measurement and the spectral index reported in $\mathsection$\ref{subsec:ATCA_obs} and \citet{2018MNRAS.475..942F} may be explained by variability over timescales of months -- years, but we caution that sub-band calibration has not been properly evaluated for our MeerKAT observations, so our measurements may be subject to $\sim 10$\,per\,cent systematic errors. 

To check short-timescale variability, we imaged the target with an integration time of 1 minute which resulted in 40 images for each observation. The lightcurves showed a relatively low modulation index (flux density standard deviation divided by the mean) of $\sim20$\,per\,cent and had a reduced $\chi^2$ of 1.4 for 79 degrees-of-freedom (80 observations minus one parameter for the mean). There was no evidence for minute-scale variability.
The fractional circular polarisation was $\sim9$\,per\,cent, which was broadly consistent with ASKAP measurements.
To search for linearly polarized emission from the source, we split the whole bandwidth in 16 parts (53.5\,MHz bandwidth for each part). No linearly polarized emission was detected above the detection threshold (with a $5\sigma$ upper limit of 0.1\,mJy\,beam$^{-1}$) with a maximum accessible rotation measure $\vert{\rm RM}\vert \lesssim 4000$\,rad\,m$^{-2}$.

We also produced dynamic spectra for MeerKAT observations (with the same procedure as that for ASKAP observations, see $\mathsection$\ref{subsec:ASKAP_obs}). 
There was no evidence of variability beyond the eclipsing behaviour.



\subsection{MeerKAT pulsar search}
\label{subsec:mkt_psrsearch}
We carried out a periodicity search of \blinkyaskap{} using high time-resolution pulsar search observations with MeerKAT, recorded with the PTUSE backend simultaneously with the imaging observations described above.

To account for possible variability in the pulse properties (e.g. in the scattering timescale), we conducted a periodicity search on 20-minute time-spans taken from observations on the 7th, 9th, 19th, and 20th of March 2022, using the full array at L-band. We divided the data in to two sub-bands (856-1284 MHz; 1284-1712 MHz) and searched each sub-band separately using {\sc peasoup}\footnote{\url{https://github.com/ewanbarr/peasoup}}
which is a GPU implementation of a dedispersion and time domain resampling acceleration search algorithm. We used a dispersion measure (DM) range of up to $1500 \rm \, pc \, cm^{-3}$ (chosen to span all plausible Galactic DMs based on the YMW16 model \citealp{2017ApJ...835...29Y}) and spanned an acceleration search range of $|50| \, \rm m \, s^{-2}$, which is particularly sensitive for detecting circular binary systems with a maximum companion of mass 0.5\,$M_{\odot}$. The candidates obtained after the search were folded following the methodology outlined in \citet{2023MNRAS.522.1071S} before undergoing manual scrutiny. No convincing candidates were found above a signal-to-noise ratio (S/N) of 8. We also repeated the same methodology for the full frequency band, ranging from 856 MHz to 1217 MHz. However, this analysis also yielded no convincing candidates.

The four observations were also independently searched, within the same DM range, with \textsc{pulsar\_miner} \citep{2021MNRAS.504.1407R}, an automated pulsar searching pipeline based on \textsc{presto}\footnote{\url{https://github.com/scottransom/presto}} \citep{Ransom_New_search_techniques_2001}. With the latter, we performed acceleration searches in the Fourier domain using the \textsc{accelsearch} routine, with a maximum allowed Doppler-induced Fourier bin drift $z_{\rm max} = \pm 200$. Besides being searched in its full 20 minutes, each observation was also split into two segments of 10 minutes, and searched individually, but no splitting of the observing band was made. Similarly to {\sc peasoup}, \textsc{pulsar\_miner} returned no convincing pulsar-like candidates.

\subsection{Murchison Widefield Array}

We obtained observations from the Galactic Plane Monitoring program (GPM; project code G0080) undertaken at 185--215\,MHz with the Murchison Widefield Array \citep[in ``Phase \textsc{ii} extended-configuration''][]{2013PASA...30....7T,2018PASA...35...33W} in June to September 2022. The survey covered $|b|<15^\circ$ and $284^\circ<l<73^\circ$ using 10$\times$30-min pointings repeated at a cadence of approximately $3$ -- $4$\,d. The survey will be described in full by Hurley-Walker et al., \textit{in prep.}. One output of the survey are flux-density and position-calibrated multi-frequency-synthesis 5-minute snapshot images centred at 200\,MHz. \blinkyaskap{} lies in the most sensitive part of the main lobe for 30\,min per night, and in adjacent pointings for a further 30\,min either side, at decreased sensitivity. We used the \textsc{aegean} source-finder \citep{2012MNRAS.422.1812H,2018PASA...35...11H} to measure the flux density at the location of \blinkyaskap{} in these images, using the priorised-fit mode, in which the position is fixed and the flux density is fitted (this obtains unbiased estimates when the source is off). The time of measurement was set to the middle of the 5-minute integration\footnote{While finer timescale ($\sim4$\,s) data is available, noise levels of $\sim60--300$\,mJy\,beam$^{-1}$ (depending on the pointing) render the signal-to-noise too low for detailed measurements of the ingress and egress.}.

\begin{figure}
    \centering
    \includegraphics[width=\linewidth]{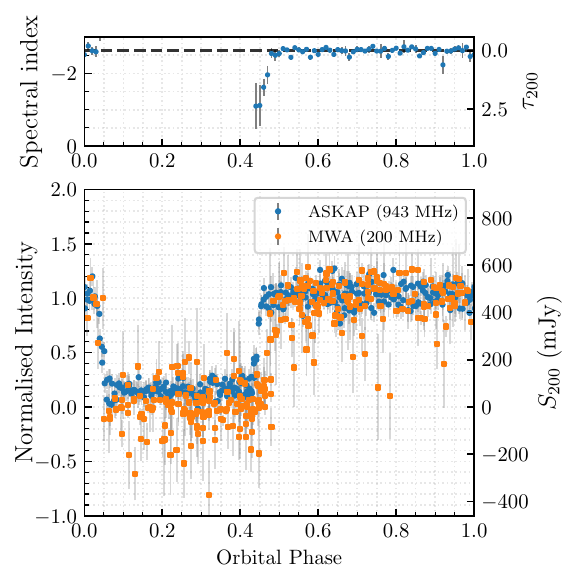}
    \caption{Normalized flux density light-curves over orbital phase from ASKAP and MWA observations. The top panel shows the inferred spectral index. On the right-hand axes we show the inferred 200\,MHz optical depth $\tau_{200}$ and 200\,MHz flux density ($S_{200}$) respectively. {The apparent discrepancy in the eclipse depth between ASKAP and MWA is the result of forced fit for flux density measurements in ASKAP observations, resulting in small but positive flux densities during eclipse when the source is not detected.}}
    \label{fig:mwa_askap_spidx}
\end{figure}

In Fig. \ref{fig:mwa_askap_spidx}, we show the folded, normalized MWA flux density time-series at the period of \periodhval{}\,h derived from ASKAP observations, which shows similar eclipse-like behaviour as described in $\mathsection$\ref{subsec:ASKAP_obs} (see Fig. \ref{fig:eclipse_lc}). We discuss these eclipses further in \S\ref{subsec:eclipse}.

We also retrieved archival measurements from the MWA Transients Survey \citep[MWATS][]{2019MNRAS.482.2484B}, which was carried out using an earlier, more compact configuration of the MWA. We observed similar eclipse-like variability to that described above, albeit with lower S/N owing to the lower angular resolution. Due to the lower quality of data, we do not consider these observations any further in this paper.

\subsection{Archival infra-red, optical, and X-ray observations}



We searched for archival optical and infra-red observations coincident at the source location. The deepest available observations were from the Vista Variables in the Via Lactea \citep[VVV;][]{2010NewA...15..433M} in infrared and the Dark Energy Camera Plane Survey  \citep[DECaPS; ][]{2023ApJS..264...28S} in optical wavelengths.

Inspection of archival infrared images from VVV and DECaPS shows a $H\sim21$ infrared source spatially coincident with \blinkyaskap{}, located at $16^{\rm h}46^{\rm m}22^{\rm s}.74$, $-44\degr05\arcmin41\farcs1$. The separation between the infrared source position and radio source position is $0\farcs12$ --  within the astrometric uncertainty of the radio position. The source is also detected in $J$-band ($1.2\,\mu$m), but diffraction spikes from a nearby bright ($J\sim 10$) star make accurate flux measurements challenging. The putative counterpart is also detected in DECaPS in the $g$, $r$, and $z$ bands. We provide the detection AB magnitudes in Table \ref{tab:count_tab}. For DECaPS, we computed the false-alarm rates based on the density of sources with $r$-band magnitudes brighter than the putative counterpart within 0.1\,deg, and find rates of $< 10^{-3}$ for each epoch.  This suggests that the source identified here is indeed the counterpart to \blinkyaskap{}.  The non-detection in VVV Z-band but detection in DECaPS $z$ band is puzzling, but may be due to intrinsic variability from the companion as seen in other spider-type systems \citep[e.g.,][]{2019ApJ...883..108D}. We discuss the nature of the optical/infra-red source further in \S \ref{subsec:oir_nature}. 

{We retrieved 4.01\,ks of archival X-ray observations from the \textit{Swift X-ray Telescope} (\textit{XRT}) aboard the \textit{Neils Gehrels Swift Observatory} \citep{2005SSRv..120..165B}. We merged the individual observations with the online analysis tools\footnote{\url{https://www.swift.ac.uk/user_objects/}}. There were 2 counts within 15\arcsec of \blinkyaskap{}, but this is consistent with the background. Using a Monte-Carlo method by summing counts in independent randomly-placed PSF-sized regions across the image (excluding positions of catalogued X-ray sources), we set a 95\,per\,cent upper limit of $7.5\times 10^{-4}$\,s$^{-1}$ in the 0.3--10 keV band.
We estimated the upper limit of H\,I column density for the position of \blinkyaskap{} based on the HI4PI survey \citep{2016A&A...594A.116H} using the HEASARC web-based \textsc{nH} tool\footnote{\url{https://heasarc.gsfc.nasa.gov/cgi-bin/Tools/w3nh/w3nh.pl}} to be $1.42\times 10^{22}$\,cm$^{-2}$.
Assuming a power-law photon index of $\Gamma=2.0$ \citep[e.g.][]{2018ApJ...864...23L}, we obtain an upper limit on the unabsorbed flux (0.3--10\,keV) of $7.6\times 10^{-14}$\,erg\,cm$^{-2}$\,s using the HEASARC web-based \textsc{PIMMS} tool\footnote{\url{https://heasarc.gsfc.nasa.gov/cgi-bin/Tools/w3pimms/w3pimms.pl}}}.

\begin{table}
    \centering
    \begin{tabular}{ccc}
    \hline
    {Filter} &
    {$\lambda_{\rm eff}$} &
    {${\rm AB}_{\rm mag}$}~$^\text{\textit{a}}$ \\
    {} &
    {($\mu$m)} &
    {(mag)}\\
    \hline \hline
    DECaPS $g$ & $0.480$ & $23.56 \pm 0.18$ \\
    DECaPS $r$ & 0.638 & $22.09 \pm 0.07$ \\
    DECaPS $i$ & 0.777 & $>22.24$ \\
    DECaPS $z$ & 0.911 & $21.8 \pm 0.2$\\
    DECaPS $Y$ & 0.985 & $>21.1$\\
    VVV Z & 0.878 & $>23.20$ \\
    VVV Y & 1.021 & $21.13\pm0.26$ \\
    VVV J & 1.254 & $19.20\pm0.09$~$^\text{\textit{b}}$ \\
    VVV H & 1.646 & $20.94\pm0.28$ \\
    VVV ${\rm K}_{\rm s}$ & 2.149 & $20.68\pm0.32$\\
    \hline
    \end{tabular}
    \caption{AB magnitudes of the possible counterpart to \blinkyaskap{}. Notes -- \textit{a}: we report $5\sigma$ magnitude lower limits for non-detections; \textit{b}: this measurement is affected by diffraction spikes from a nearby bright star.\label{tab:count_tab}}
\end{table}








\section{Discussion}\label{sec:discussion}

\subsection{The nature of the source}

We have discovered periodic ($P_b=~$\periodh{}\,h, see \S\ref{subsec:ecl_props}), eclipse-like radio variability from \blinkyaskap{} spatially coincident with the $\gamma$-ray source \blinky{}. The $\gamma$-ray source is offset by $95\farcs 2$ at a position angle of $290^\circ$ from the radio source, well within the \textit{Fermi} positional error ellipse (see Fig. \ref{fig:field_image}). The radio source is point-like, exhibits a steep spectral index ($\alpha\sim -2.7$), and has significant circular polarisation ($f_c\sim 7$\,per\,cent). Furthermore, we have identified a potential faint ($g \sim 23.5$, $H\sim 21$) and variable counterpart using archival optical and infra-red observations.  

The only known source class that satisfies this set of characteristics are spider MSP binary systems. In particular, in \S\ref{subsec:eclipse}, we infer an eclipse duration of $0.3994\pm{0.0014}$ orbits at $943$\,MHz. The system is therefore most likely a ``redback'', which typically exhibit longer eclipses ($\sim 50$\,per\,cent of the orbit; \citealp{2021ApJ...909....6D,2016ApJ...819...34C}) as opposed to black widows that typically eclipse for $\sim 10$\,per\,cent of the orbit \citep[e.g.][]{2020MNRAS.494.2948P}. This is consistent with the higher mass, main-sequence (non-degenerate) companions in redback systems, which are capable of producing larger eclipse regions as the pulsar ablates the companion atmosphere. Based on these insights, we claim that the radio source \blinkyaskap{} is associated with the $\gamma$-ray source \blinky{}, and refer to it as \blinky{} from hereon. Future detection of pulsations in radio and $\gamma$-rays would confirm the association.

{Multiple spider systems have been previously detected at X-ray wavelengths  with redbacks in particular exhibiting relatively high X-ray luminosities around $10^{31}$ -- $10^{32}$\,erg\,s \citep{2022MNRAS.511.5964Z,2018ApJ...864...23L}. Our unabsorbed X-ray flux upper limit of $<7.6\times 10^{-14}$\,erg\,cm$^{-2}$\,s implies a luminosity upper limit of $<9\times 10^{30}$\,erg\,s$^{-1}$\,$(d / 1\,\text{kpc})^{2}$. The X-ray non-detection could therefore be explained if the distance is greater than 1\,kpc, although this lower limit estimate is subject to uncertainties around the true hydrogen column density and scatter in the redback X-ray luminosity distribution.}

We compiled integral $\gamma$-ray fluxes above 100\,MeV ($E_{100}$), radio flux densities ($S_\nu$), along with DMs (where available) for known spider-type binaries using the ATNF Pulsar Catalogue \citep{2005AJ....129.1993M} and the {Third \textit{Fermi}-LAT Catalogue of $\gamma$-ray Pulsars \citep[][hereafter \fermi{} 3PC]{2023arXiv230711132S}}.

\begin{figure*}
    \centering
    \includegraphics[width=\linewidth]{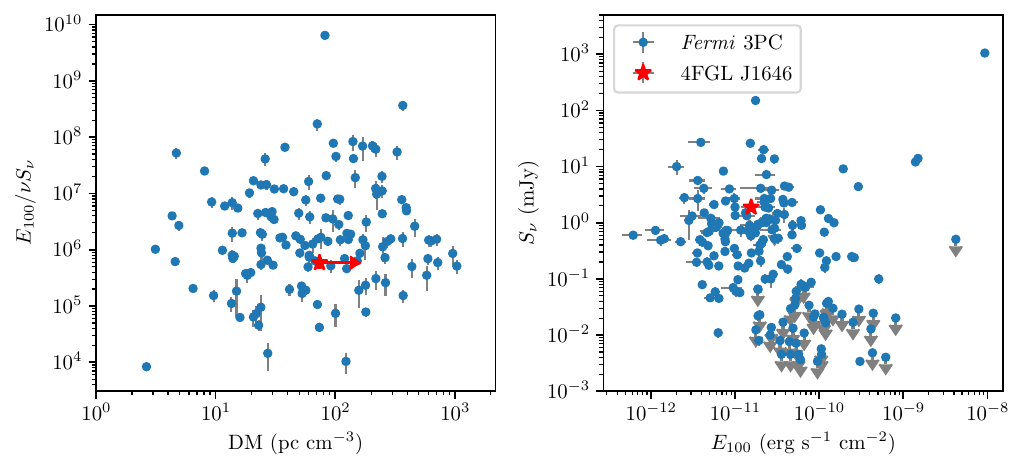}
    \caption{Radio and $\gamma$-ray emission properties from the {\fermi\ 3PC and for \blinky}. Left: Ratio of $E_{100}$ to $\nu S_\nu$ as a function of pulsar DM. Right: $S_\nu$ versus $E_{100}$. \blinky{} is marked with a red star in both panels. {The DM lower limit of $\sim74$\,pc\,cm$^{-3}$ for \blinky\ is derived from scattering properties derived from the NE2001 Galactic electron density model \citep{2002astro.ph..7156C}}, based on the estimated DM at which diffractive interstellar scintillation becomes undetectable in our observations. The location of \blinky{} in this parameter space is consistent with the \fermi{} $\gamma$-ray pulsar population.}
    \label{fig:gammaray-radio-paramspace}
\end{figure*}

In Fig.~\ref{fig:gammaray-radio-paramspace} (left panel), we show the ratio of $E_{100}/\nu S_\nu$ versus DM. This quantity, representing the relative flux densities of $\gamma$-ray emission to radio emission, is independent of distance, and therefore by proxy is independent of DM. We computed $E_{100}/\nu S_\nu$ for \blinky{}, and adopted a lower limit on the DM of 74\,pc\,cm$^{-3}$ (as estimated in $\mathsection$\ref{subsec:stableemission}) to place \blinky{} in this parameter space. As can be seen, the properties of \blinky{} are consistent with the population of radio- and $\gamma$-ray detected spider binaries. 

In Fig.~\ref{fig:gammaray-radio-paramspace} (right panel), we show $S_\nu$ as a function of $E_{100}$, for pulsars in \fermi{} 3PC and for \blinky{}. Once again, the parameters of \blinky{} are consistent with the radio-loud population of $\gamma$-ray pulsars.

In the following subsections, we discuss further constraints on the nature of the system with the available observations.

\subsection{Nature of the optical/infra-red companion}
\label{subsec:oir_nature}
\begin{figure}
\centering
    \includegraphics[width=\linewidth]{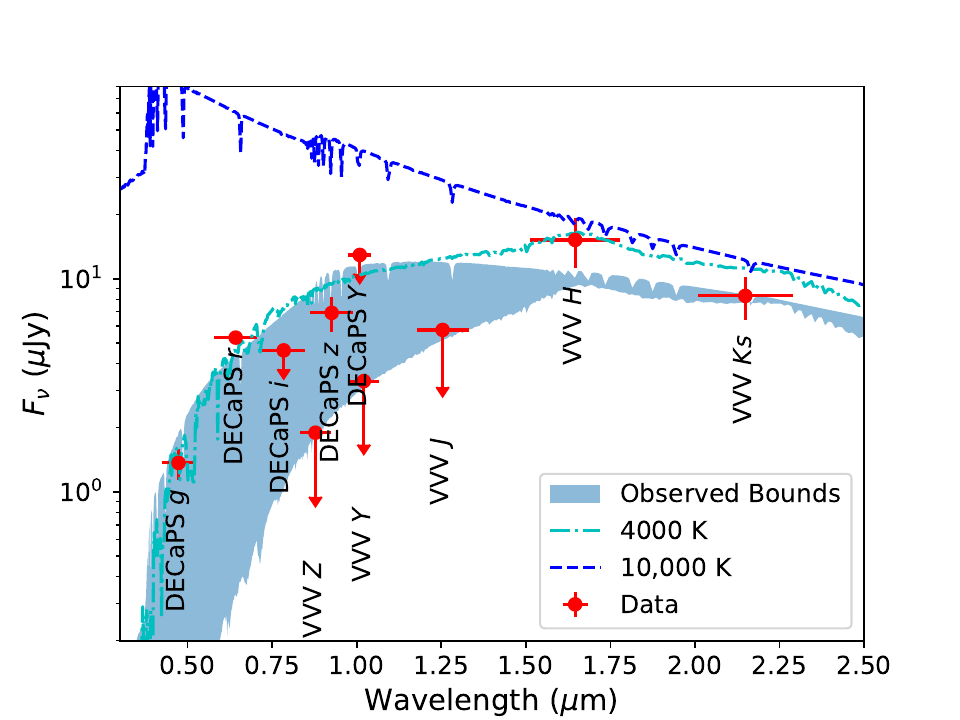}
    \caption{Spectral energy distribution of the optical/IR counterpart to \blinky{}.  We show the flux densities from VVV and DECaPS (Table~\ref{tab:count_tab}).  We also show model atmospheres from \citet{2003IAUS..210P.A20C}.  We include atmospheres for effective temperatures of $10,000$\,K (blue dashed line) and 4000\,K (cyan dot-dashed line), roughly bounding the observed range for many spiders.  The models are normalized to a radius 50\,per\,cent of the Roche lobe radius and a distance of {3\,kpc}. These models, with an extinction of {$A_V=3.4$\,mag} applied are then shown in the blue shaded region as the observed bounds, which roughly bound the detections and upper limits of the photometry.}
    \label{fig:oir}
\end{figure}

We show the available optical/infrared photometry for the counterpart to \blinky{} in Fig.~\ref{fig:oir}.  Note that while VVV and DECaPS both have multiple epochs (and so should allow for variability searches), the crowded field and marginal nature of some of the detections meant that we were limited to a single average detection for each.  As noted above, the data from VVV and DECaPS are not consistent with each other: the source is detected in DECaPS $z$ band but not VVV $Z$ band, with contradictory upper limits.  However, many spider systems have significant optical variability from a combination of ellipsoidal modulation and irradiation \citep[e.g.,][]{2019ApJ...872...42S,2019ApJ...883..108D,2023MNRAS.520.2217M}.  While we do not have sufficient data to do a complete solution of this lightcurve, we can at least see if the available data are plausibly fit by a variable companion to an energetic pulsar.

We assume that the companion has a radius of 50\,per\,cent of the Roche lobe radius (on the small end but consistent with those from \citealt{2023MNRAS.520.2217M}) at a distance of 3\,kpc.  We take effective temperatures of 4000\,K and 10,000\,K (again, reasonably consistent with those of the population, although some non-irradiated faces can be significantly cooler).  After accounting for an extinction $A_V=3.4\,$mag, we find that the predicted range in flux density is consistent with our observed photometry.  This is not a unique solution, but just a plausibility argument: the Roche lobe filling factor and distance are entirely degenerate, and the effective temperatures can also be varied by quite a bit.  Nonetheless we see that the optical/IR emission from the counterpart to \blinky{} does not require an emission region larger than the Roche lobe or an effective temperature much hotter or cooler than other similar sources.  Future time-resolved photometry can help establish a more unique solution.

\subsection{Implications from non-eclipsed radio emission}
\label{subsec:stableemission}
The lack of significant variability within non-eclipsed time periods in any of our observations implies the scintillation timescale or bandwidth is lower than the dynamic spectrum resolution (temporal resolution $\delta_t = 8\,{\rm s}$ and frequency resolution $\delta_\nu = 0.84\,{\rm MHz}$ for MeerKAT).

Using this non-detection of scintillation, we can place an approximate lower limit on the dispersion measure of $\sim 74\,{\rm pc}\,{\rm cm}^{-3}$, using the NE2001 Galactic electron density model \citep{2002astro.ph..7156C}, under the condition that the scintillation bandwidth $\Delta \nu_d < 2\delta_\nu$. 

In addition, we can also rule out putative rotation periods exceeding $2\delta_t=16$\,s. While this is not particularly informative under the spider MSP scenario for \blinky{}, it does rule out \blinky{} belonging to the emerging class of recently-discovered ultra-long period objects \citep{2022Natur.601..526H,2022NatAs...6..828C,2023Natur.619..487H} or other more unusual radio variables such as GCRT-like sources \citep{2021ApJ...920...45W,2005Natur.434...50H}.

We also observe long-term variability in the non-eclipsed flux density, as can be seen in Fig. \ref{fig:eclipse_lc}. This is typical of spider systems \citep{2023ApJ...942...87K,2020MNRAS.494.2948P}

\subsection{Non-detection of radio pulsations}

\begin{figure}
    \centering
    \includegraphics[width=\linewidth]{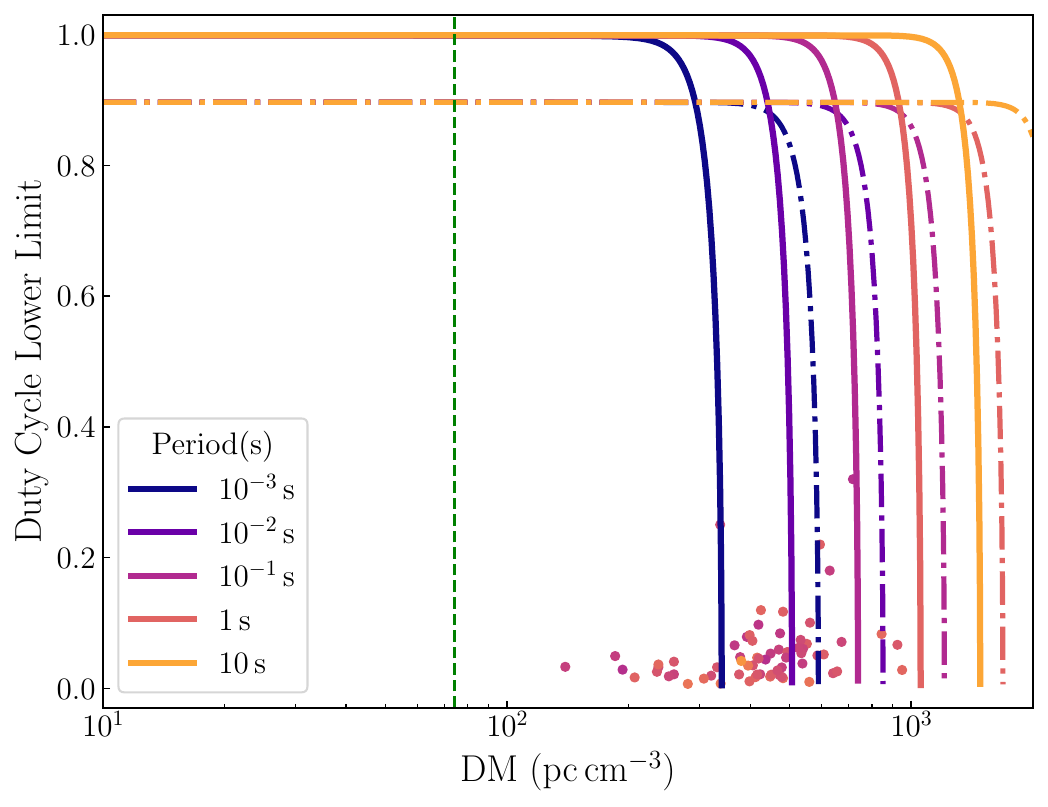}
    \caption{Duty cycle constraints as a function of dispersion measure, for a range of pulsar periods. Coloured points show pulsars from the ATNF Pulsar Catalogue \citep{2005AJ....129.1993M} within $4^\circ$ of \blinky. These pulsars all lie below and to the left of the DM-duty cycle locus that defines the boundary of this parameter space where they are detectable, meaning that their pulsations are detectable. 
    Solid lines are the limits derived from the MeerKAT pulsar searches based on the full bandwidth (856--1217\,MHz) data, and dash-dotted lines are the limits derived from the Murriyang UWL pulsar searches based on the top four subbands (3520--4032\,MHz) data (expected flux density was derived based on the spectral index from MWA and ASKAP observations $\alpha=-2.7$).
    }
    \label{fig:dutycycle}
\end{figure}

Our search for pulsed radio emission searched from Murriyang and MeerKAT observations yielded no viable candidates.
A likely scenario is that the pulses are smeared due to high degrees of scattering either from the intra-binary or interstellar material. In Fig. \ref{fig:dutycycle}, we show the detectability thresholds for intrinsic pulse duty cycle lower limits as a function of DM for a range of spin periods, owing to scattering in the interstellar medium, using the scattering timescale-DM relationship from \citet{2004ApJ...605..759B}. At the DM lower limit of 74\,pc\,cm$^{-3}$, our lack of detected pulsations implies a pulsar-intrinsic duty cycle close to 1, which we consider unlikely. Instead, an interstellar DM exceeding $\sim 600$\,pc\,cm$^{-3}$ would rule out virtually all millisecond pulsars from detectability. On the other hand, the scattering may (partially) arise from the intra-binary material, in which case the DM-scattering relationship is less applicable. We discuss this further in \S \ref{subsec:eclipse}.

Alternatively, the lack of radio pulsations could arise from an unusual emission geometry. Typically, spiders are detected with radio and $\gamma$-ray pulsations, but some are also radio-quiet \citep{2012ApJ...747L...3K,2022ApJ...935....2C}. This radio-loud/quiet dichotomy is not fully understood, but it may be due to eclipse phenomena, or geometry of the radio vs. $\gamma$-ray beam. Here, we have an intermediate case: radio emission, $\gamma$-ray emission, but an apparent lack of radio pulsations. While the putative MSP likely produces $\gamma$-ray pulses, we did not attempt to search for them due to the excessive computational demands without constrained orbital elements. If the geometry of the system can be constrained by other means (e.g. through optical monitoring or through detection and timing of $\gamma$-ray pulsations), this system may provide key insights into the apparent dichotomy between radio-loud and quiet spider binaries, and/or the geometry of MSP emission.



\subsection{Derived eclipse properties}
\label{subsec:ecl_props}

The phenomenology of eclipses carries information on the nature of the system and the properties of the intra-binary material \citep{2020MNRAS.494.2948P}. To determine the eclipse properties, we fitted the ASKAP and MWA lightcurves with a double Fermi-Dirac function;
\begin{equation}
\label{eq:fermidirac}
\begin{split}
    \frac{S_\nu (\phi,t)}{S_{\text{nonecl}}(t)} =   \beta \left( \frac{1}{\exp\left({\frac{(\phi - \phi_i)}{w_i}}\right) + 1} - \frac{1}{\exp\left({\frac{(\phi - \phi_e)}{w_e}}\right) + 1}\right)  \\ 
    + (1 - \beta),
    \end{split}
\end{equation}
where $\phi \equiv \frac{(t - T_0)\%  P_b }{P_b}$ is the derived orbital phase\footnote{The symbol "$\%$" is the modulo operator.}, where $t$ is the MJD time, $P_b$ and $T_0$ are the binary period and reference MJD epoch\footnote{Here we define the reference epoch such that the mid-point of the eclipse, which approximates superior conjunction of the companion, falls at orbital phase 0.25, following the pulsar convention. Note that because we do cannot derive a full binary solution, $T_0$ in our model is likely only an approximation to the epoch of ascending node.}, which we allow to vary. Other fitted parameters include $S_\mathrm{nonecl}(t)$, the non-eclipsed peak flux density, the ingress/egress phase $\phi_i$ and $\phi_e$, and the $1/e$-width of the ingress/egress, $w_i$ and $w_e$. As can be seen in Fig.~\ref{fig:eclipse_lc}, the non-eclipsed flux density shows substantial long-term variability at 943\,MHz. To account for this, we modelled $S_\mathrm{nonecl}$ as a third-order polynomial in $(t-T_0)$, allowing the polynomial coefficients to vary in the fit.

The constraints on $P_b$ and $T_0$ from MWA measurements were less precise than ASKAP constraints by about an order of magnitude. To ensure consistency between the measured eclipse properties between ASKAP and MWA data, we used a multivariate normal prior on $P_b$ and $T_0$ for fitting MWA data, with the mean and covariance of the prior derived from the ASKAP marginal posteriors. We used uniform priors on all other parameters for both MWA and ASKAP fits.

Using the broad frequency coverage afforded by ASKAP and MWA, we measured the frequency dependence of the eclipse properties parametrised with a power-law spectral index, e.g. $\Delta \phi \propto \nu^{\alpha_{\Delta \phi}}$. 
We provide the properties in Table \ref{tab:eclipse_table}, along with the derived eclipse radius $R_E = \pi a \Delta \phi$, where $a$ is the semimajor axis assuming a $1.4\,M_\odot$ pulsar and a companion mass of either $0.4\,M_\odot$ corresponding to typical redback companion masses \citep{2019ApJ...872...42S}.

\begin{table*}
    \centering
    \begin{tabular}{cccc}
    \hline
    {Eclipse parameter}  & {MWA} & {ASKAP} & {Spectral index} \\
    {}        & {(200\,MHz)}                      & {(943\,MHz)}      & {}\\
    \hline
    \hline
    $P_b$\,(h)           & $-$                       & \periodh{} & \\
    $T_0$\,(MJD)         & $-$                       & $59391.3567\pm0.0004$ & \\
    $\phi_i$             & $0.047^{+0.003}_{-0.007}$ & $0.0502\pm0.0016$ & $0.003^{+0.005}_{-0.003}$\\
    $w_i$                & $< 0.0064$\,$^\text{\textit{a}}$  & $0.0101\pm0.0008$                   & $1.2^{+1.1}_{-0.8}$ \\
    $\phi_e$             & $0.491\pm 0.002$          & $0.4497\pm0.0016$           & $-0.121\pm0.008$ \\
    $w_e$                & $0.0118^{+0.0013}_{-0.0014}$ & $0.0089\pm0.0009$           & $-0.2\pm0.1$\\
    $\Delta \phi$        & $0.444^{+0.007}_{-0.003}$ & $0.3994\pm0.0014$   & $-0.069^{+0.005}_{-0.009}$\\
    $R_E$ $^\text{\textit{b}}$ & $2.3\,R_\odot$             & $2.6\,R_\odot$ & \\
    \hline
    \end{tabular}
    \caption{Properties of the radio eclipses derived from fitting a Fermi-Dirac function as in Equation \ref{eq:fermidirac}. The spectral index is defined such that the quantity in each row follows a power-law, e.g. $\Delta \phi \propto \nu^{\alpha_{\Delta \phi}}$. Notes (\textit{a}): 90\% confidence upper-limit; (\textit{b}): {assuming a companion mass of $0.4\,M_\odot$}.\label{tab:eclipse_table}}
\end{table*}

The inferred eclipse radius is $\sim2.3\,R_\odot$, which is $\sim1.5$ times larger than the Roche lobe radius $R_L\sim 1.6\,R_\odot$ assuming a pulsar mass of $1.4\,M_\odot$ and companion mass of $0.4\,M_\odot$ \citep[computed using the approximation of][]{1983ApJ...268..368E}, confirming that the eclipses are due to an unbound cloud of material from the companion. The source of this material could either be the ablated material from the companion, and/or the wind from the putative pulsar itself.

\subsection{Nature of the radio eclipses}
\label{subsec:eclipse}
The phenomenology of radio eclipses in spider systems is very broad, considering eclipses both in individuals systems and across different systems \citep{2019MNRAS.490..889P, 2020MNRAS.494.2948P}. Multiple (possibly overlapping) mechanisms have been suggested to drive the eclipses -- see e.g. \citet{1994ApJ...422..304T} for analytic treatment of eclipse mechanisms and \citet{2020MNRAS.494.2948P} for an overview of observational results across multiple systems. The eclipse  mechanisms may influence the pulse properties (e.g., DM or scattering), and can also absorb the continuum flux density. In the case of \blinky{}, we observe 100\,per\,cent absorption of the continuum flux density, but the lack of observed pulsations renders other observables (e.g. change in DM) unattainable.

Nonetheless, our relatively long observational time-span from ASKAP and MWA observations, and the broad frequency coverage they afford, enable us to investigate the properties of the eclipse and make some initial estimates of the physical conditions within the eclipse medium.

Following similar arguments to e.g. \citet{2021ApJ...920...58K}, we investigated eclipse mechanisms proposed by \citet{1994ApJ...422..304T}: cyclotron absorption, induced Compton scattering, free-free absorption, and synchrotron absorption. Without strong constraints on the properties of the companion or the binary orbit, we can only make broad estimates of the properties of the eclipse medium. Nonetheless, we carry out this exercise to determine if any eclipse mechanisms can be ruled out and to estimate a plausible range of eclipse medium properties. For the following calculations we assume a companion mass of $0.4\,M_\odot$, which is typical for redback systems \citep{2019ApJ...872...42S}, and a pulsar mass of $1.4\,M_\odot$. The measured period of $0.2194594\,$d implies an orbital semi-major axis of $\sim 1.9\,R_\odot$ for the assumed companion mass.

For cyclotron absorption, we estimate the eclipse region magnetic field assuming equilibrium between the wind energy density $U_E = \dot{E}/4\pi c a^{2}$ and the magnetic pressure $P_B = B_E^2 / 8\pi$. Assuming $\dot{E}\sim 10^{34}$\,erg\,s$^{-1}$, we have $U_E \sim 1.6$\,erg\,cm$^{-3}$, and $B_E\sim 6.3$\,G. We find that the required temperature to produce eclipses at 943\,MHz is $>4\times10^7$\,K, which exceeds the $1.7\times10^{6}$\,K upper threshold for cyclotron absorption \citep{1994ApJ...422..304T}. We therefore rule it out as the eclipse mechanism.

For induced Compton scattering, we varied the electron column density (between $10^4$ and $10^{30}$\,cm$^{-2}$), temperature (between $10^2$ and $10^9$\,K), and de-magnification factor (between $0$ and $1$) due to reflection of the radiation off a curved plasma cloud. The maximum optical depth we calculated across the parameter space was $\tau \sim 10^{-2}$, so we therefore rule out induced Compton scattering as the eclipse mechanism.

To determine the viable range of parameters for free-free and synchrotron absorption, we calculated the optical depth for MWA 200\,MHz observations just after full eclipse egress for at 943\,MHz, around orbital phase 0.47 (i.e., when the 943\,MHz optical depth $\tau$ first returns to 0). The resulting 200\,MHz optical depth was $\tau = 2.3\pm0.9$. An important caveat for these estimates is that the measurements at the relevant orbital phase were taken several months apart, so long-term variability will introduce a systematic uncertainty of $\sim 10$\,per\,cent. Any constraints on the physical properties of the medium should therefore be taken as order-of-magnitude estimates only. To determine the physical properties, we fit the optical depth for free-free and synchrotron absorption using Markov Chain Monte Carlo, fitting to the 943\,MHz and 200\,MHz flux density at the chosen orbital phase $\phi = 0.47$ during partial egress.

For free-free absorption, we varied the electron column density between $10^{10}\,\mathrm{cm}^{-2}$ to $10^{25}\,\mathrm{cm}^{-2}$ -- a very broad range around typical column densities of $\sim 10^{16}\,\mathrm{cm}^{-2}$ \citep[e.g.][]{1996ApJ...465L.119S,2022MNRAS.513.1794B}. We also allowed the temperature to vary between $10^0$\,--\,$10^{10}$\,K, and the clumping factor to vary between $10^0$\,--\,$10^{10}$. We held the size of the absorbing medium fixed at $2R_E = 4.7\,R_\odot$ assuming a $0.4\,M_\odot$ companion. 

Synchrotron absorption is caused by a population of non-thermal electrons, often assumed to possess a power-law energy distribution $n(E) = n_0 E^{-p}$ between minimum and maximum energies $E_\text{min}$ and $E_\text{max}$ \citep{1994ApJ...422..304T}. We allowed the reference volume density of non-thermal electrons $n_0$ to vary between $10^1\,\text{cm}^{-3}$ to $10^{8}\,\text{cm}^{-3}$, the eclipse average magnetic field strength between $10^{-2}$\,G to $10^2$\,G, and the electron energy power-law index $p$ between 0 and 8. We held the viewing angle $\theta$ fixed to $\pi/3$ (corresponding to the expectation value of $\theta$ assuming a uniform distribution in $\cos \theta$). As with free-free absorption, we held the size of the absorbing medium fixed at $2R_E = 4.7\,R_\odot$.

\begin{figure}
    \includegraphics[width=\linewidth]{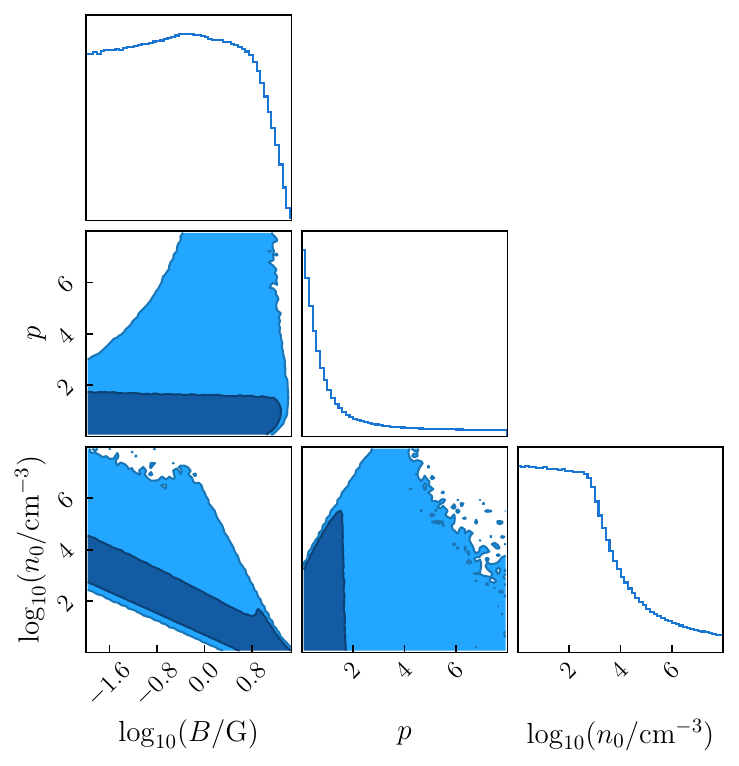}
    \caption{Marginal posterior probability densities of eclipse medium properties (non-thermal electron density $n_0$, magnetic field strength $B$, and electron energy power-law index $p$) assuming synchrotron absorption. \label{fig:tau_synch}}
\end{figure}

\begin{figure}
    \includegraphics[width=\linewidth]{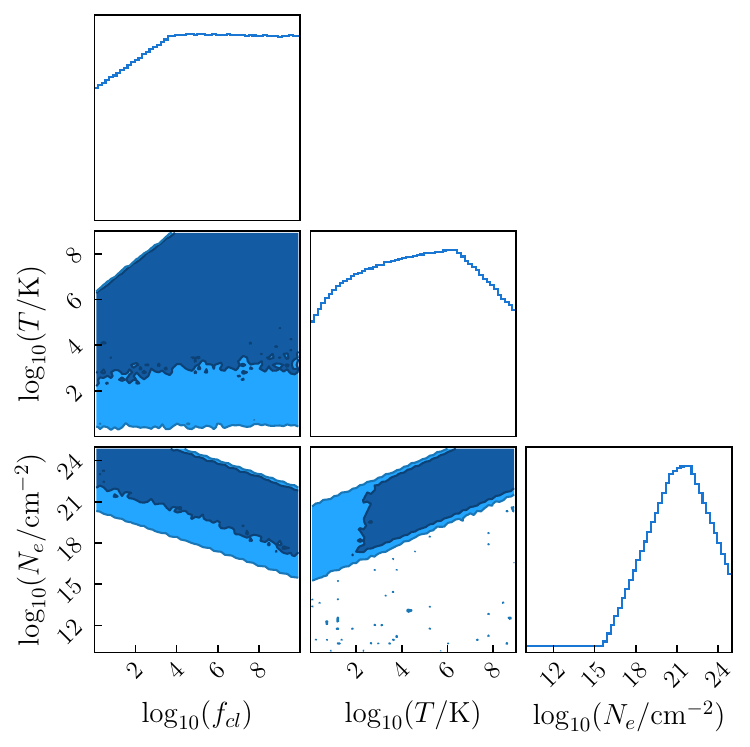}
    \caption{As with Fig. \ref{fig:tau_synch}, but for eclipse medium properties (clumping factor $f_{\mathrm{cl}}$, temperature $T$, and electron column density $N_e$) assuming free-free absorption. \label{fig:tau_ff}}
\end{figure}

Figs \ref{fig:tau_synch} and \ref{fig:tau_ff} show the marginal posterior probability densities on each of the varied parameters in the synchrotron and free-free absorption models. We found that a broad range of parameters could reproduce the observed optical depths at the relevant orbital phase, but we are able to rule some regions of parameter space out. For instance, if free-free absorption is responsible for the eclipses, then the electron column density must exceed $10^{15}$\,cm$^{-2}$. If we assume an equipartition magnetic field strength of $\sim 6$\,G (as with cyclotron absorption) in the synchrotron-absorbing plasma, then $n_0 \lesssim 100$\,cm$^{-3}$. Independent constraints on the eclipse medium density (e.g. from measurements of DM variations near eclipse) will significantly improve constraints on the eclipse medium properties.

\citet{2020MNRAS.494.2948P} noted substantial differences in the behaviour of eclipses in the continuum and pulsed flux in a sample of black widows. These differences imply that multiple eclipse mechanisms are operating in at least some spiders -- one that absorbs the continuum flux density, and another that reduces the pulsed flux density via temporal smearing. We now explore whether a similar scenario may apply to \blinky{}, whereby the continuum flux density is absorbed by e.g. synchrotron absorption, but scattering across a more extended volume in the intra-binary material destroys the pulsed flux density across the full orbit.

\begin{figure}
    \includegraphics[width=\linewidth]{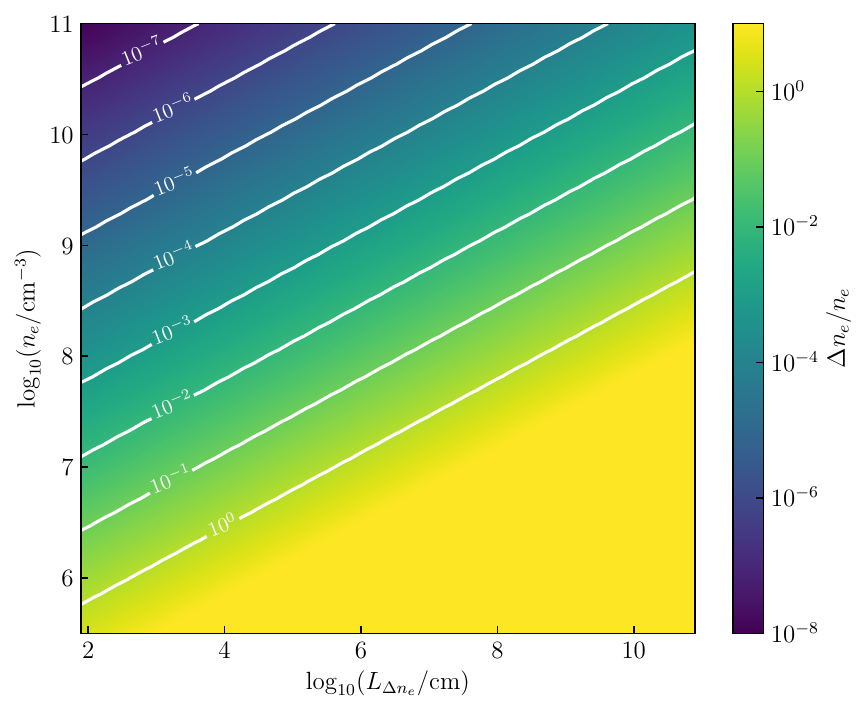}
    \caption{Constraints on the fractional electron density variations $\Delta n_e/ n_e$ required in the eclipse medium to scatter-broaden pulsed emission beyond detectability in Murriyang observations centred at 3200\,MHz, as a function of the mean electron density $n_e$ and the length-scale of density variations $L_{n}$. For clarity we show contours at each decade of $\Delta n_e / n_e$ from $1$\,--\,$10^{-7}$.\label{fig:delta_ne_scatt}}
\end{figure}

We consider an extended scattering eclipse medium with radius $R_E = 2a$, such that the pulsar is always enveloped, and its pulsed emission therefore scatter-broadened beyond detectability. \citet{1994ApJ...422..304T} considered the density fluctuations required to render pulsed emission undetectable due to scattering in a turbulent eclipse medium. The fractional density fluctuations $\Delta n_e / n_e$ over a length scale $L_{n}$ required to broaden a pulse for more than a pulsar spin period $P$ is \citep{1994ApJ...422..304T}
\begin{equation}
    \frac{\Delta n_e}{n_e} \gtrsim 7.6\times 10^{-6} \frac{\nu_9^2}{n_{e,8}}\frac{P_{-3}^{1/2}}{(R_{E,11} a_{11})^{1/2}} L_n^{1/2}\,,
\end{equation}
where the subscript numerals represent the normalization factor for the various quantities in cgs units, e.g. $n_{e,8} = n_e / 10^{8}\,\mathrm{cm}^{-3}$. We show the constraints on the density fluctuations in Fig. \ref{fig:delta_ne_scatt}. For typical densities $n_e\sim 10^{7}\,\mathrm{cm}^{-3}$ and short fluctuation length scales $L_n \sim 10^3\,\mathrm{cm}$, fractional density fluctuations of $\sim 10^{-2}$\,--\,$10^{-1}$ are required to scatter-broaden the pulses out of detectability. However, the requirements are less stringent at higher mean densities.

Better constraints on the companion and its orbit, along with ultra wide-band continuum observations will enable more precise measurements of the physical properties of the eclipse in future. If the scattering is indeed eclipse-dominated, a dedicated radio observing campaign targeting the period just before eclipse ingress -- where the density of the eclipsing material is likely to be lowest -- may yield a future detection of radio pulses.

\section{Conclusions}\label{sec:conclusion}
We have presented the discovery of radio eclipses from ASKAP~J1646$-$4406, which we associate with the $\gamma$-ray source \blinky{}. The radio-frequency and multi-wavelength properties strongly suggest that \blinky{} is a spider-type pulsar, most likely a ``redback'' with a $\sim 0.4\,M_\odot$ companion, owing to the long eclipse ($\sim 40$\,per\,cent of the orbit). We measure a putative orbital period of \periodh{}, typical for spider-type systems \citep{2019ApJ...872...42S}. Using the eclipse properties at 943 and 200\,MHz, we place broad constraints on plausible eclipse medium properties derived from synchrotron or free-free absorption models, and rule out induced Compton scattering and cyclotron absorption as the eclipse mechanism. Constraints on the electron density of the medium obtained from DM enhancements in pulsed radio emission would greatly improve the constraints on other properties of the eclipse medium. 

We searched extensively for radio pulses with Murriyang and MeerKAT, but our searches yielded no convincing pulsar candidates. The non-detection of radio pulses can be explained by temporal smearing due to angular scattering of the emission, either from propagation through the ISM or the eclipse medium. In the ISM case, the lack of pulses can be explained if the interstellar DM exceeds $\sim 600\,\mathrm{pc}\,\mathrm{cm}^{-3}$. If the scattering is dominated by the eclipse medium, a range of turbulent density fluctuations $\Delta n_e/ n_e$ could smear pulses beyond detectability.

Additional multi-wavelength studies for this source are warranted. As discussed in \S\ref{subsec:oir_nature}, available optical and infra-red measurements of the inferred counterpart plausibly agree with the hypothesis of a spider binary, but available data are not sufficient to sufficiently constrain the properties of the system. Future photometric and spectroscopic monitoring of the counterpart across the inferred orbit will be crucial both for confirming the nature of \blinky{} as a redback, and to derive a full orbital solution \citep{2020ApJ...901..156N}. 
These orbital constraints will enable a $\gamma$-ray pulsation search toward \blinky{}, which is computationally infeasible unless the orbital elements are well-determined.

\section*{acknowledgments}
Corresponding authors AZ and ZW contributed equally to this project. We thank the anonymous reviewer whose comments improved this manuscript.
Thanks to George Hobbs, Daniel Reardon and Matthew Kerr for helpful discussions. Thanks to Eddie Schlafly and Andrew Saydjari for help accessing DECaPS data.
DK is supported by NSF grant AST-1816492. RS is supported by NSF grant AST-1816904. N.H.-W. is the recipient of an Australian Research Council Future Fellowship (project number FT190100231).
Parts of this research were conducted by the Australian Research Council Centre of Excellence for Gravitational Wave Discovery (OzGrav), project number CE170100004.
AR gratefully acknowledges financial support by the research grant ``iPeska'' (P.I. Andrea Possenti) funded under the INAF national call Prin-SKA/CTA approved with the Presidential Decree 70/2016, and continuing valuable support from the Max-Planck Society.
ASKAP is part of the Australia Telescope National Facility,
which is managed by the CSIRO. Operation of ASKAP is
funded by the Australian Government with support from the
National Collaborative Research Infrastructure Strategy. ASKAP
uses the resources of the Pawsey Supercomputing Centre.
Establishment of ASKAP, Inyarrimanha Ilgari Bundara, the CSIRO Murchison Radio-astronomy
Observatory, and the Pawsey Supercomputing Centre are
initiatives of the Australian Government, with support from
the Government of Western Australia and the Science and
Industry Endowment Fund. We acknowledge the Wajarri
Yamatji as the traditional owners of Inyarrimanha Ilgari Bundara, the CSIRO Murchison Radio-astronomy
Observatory site, the Wiradjuri people as the traditional owners of the CSIRO Parkes Radio Observatory site, and the Gomeroi people as the traditional owners of the CSIRO Paul Wild Observatory site. Support for the operation of the Murchison Widefield Array is provided by the Australian Government (NCRIS), under a contract to Curtin University administered by Astronomy Australia Limited. Thanks to ATNF and SARAO Operations staff for supporting our DDT observations. 

\section*{Data Availability}
ASKAP data used in this paper can be accessed through the CSIRO ASKAP Science Data Archive (CASDA\footnote{\url{https://data.csiro.au/dap/public/casda/casdaSearch.zul}}), using the project codes and SBIDs provided.
Murriyang observations can be obtained through the CSIRO Data Access Portal (DAP)\footnote{\url{https://data.csiro.au/domain/atnf}}, and ATCA observations can be obtained from the Australia Telescope Online Archive (ATOA)\footnote{\url{https://atoa.atnf.csiro.au/}}.
MeerKAT continuum observations can be obtained from the MeerKAT archive\footnote{\url{https://apps.sarao.ac.za/katpaws/archive-search}}. Other data can be provided upon reasonable request.

\bsp	
\label{lastpage}
\end{document}